\documentclass[aps,prb,reprint,superscriptaddress,showpacs]{revtex4-1}
\usepackage{graphicx}
\usepackage{amsmath}

\begin{document}


\title{Magnetic domain-wall velocity enhancement induced by a transverse magnetic field}


\author{Jusang Yang}
\affiliation{Department of Physics, The University of Texas at Austin, Austin, Texas 78712-1081, USA}
\author{Geoffrey S. D. Beach}
\affiliation{Department of Materials Science and Engineering, Massachusetts Institute of Technology, Cambridge, Massachusetts 02139, USA}
\author{Carl Knutson}
\affiliation{Department of Physics, The University of Texas at Austin, Austin, Texas 78712-1081, USA}
\author{James L. Erskine}\email[Electronic address: ]{erskine@physics.utexas.edu}
\affiliation{Department of Physics, The University of Texas at Austin, Austin, Texas 78712-1081, USA}


\date{\today}

\begin{abstract}
Spin dynamics of field-driven domain walls (DWs) guided by Permalloy nanowires are studied by high-speed magneto-optic polarimetry and numerical simulations. DW velocities and spin configurations are determined as functions of longitudinal drive field, transverse bias field, and nanowire width. Nanowires having cross-sectional dimensions large enough to support vortex wall structures exhibit regions of drive-field strength (at zero bias field) that have enhanced DW velocity resulting from coupled vortex structures that suppress oscillatory motion. Factor of ten enhancements of the DW velocity are observed above the critical longitudinal drive-field (that marks the onset of oscillatory DW motion) when a transverse bias field is applied. Nanowires having smaller cross-sectional dimensions that support transverse wall structures also exhibit a region of higher mobility above the critical field, and similar transverse-field induced velocity enhancement but with a smaller enhancement factor. The bias-field enhancement of DW velocity is explained by numerical simulations of the spin distribution and dynamics within the propagating DW that reveal dynamic stabilization of coupled vortex structures and suppression of oscillatory motion in the nanowire conduit resulting in uniform DW motion at high speed.
\end{abstract}


\maketitle

\section{\label{sec:1}Introduction}
A variety of recently proposed spintronic devices would use magnetic domain walls (DWs) confined by nanometer-scale magnetic conduits to store and process information.\cite{Allwood2002,Allwood2005,Parkin2008,Hayashi2008,Xu2008} Similar magnetic structures have been proposed for lab-on-a-chip platforms for manipulating and functionalizing magnetic nanoparticles, captured by DW stray fields.\cite{Vieira2009,Lee2004,Vavassori2010} The viability of DW-based spintronic technologies rests largely on how fast DWs can be propelled in nanoscale structures. It has recently been experimentally demonstrated\cite{Beach2005} that DWs driven by moderate strength longitudinal magnetic fields along a ferromagnetic nanowire obey a \textquotedblleft\textit{speed limit}.\textquotedblright  Wall velocity increases with increasing drive field but above a critical field, the \textquotedblleft\textit{harder}\textquotedblright the DW is pushed, the more slowly it moves. This counterintuitive behavior results from the nucleation and gyrotropic motion of vortices within a DW, resulting in oscillatory DW displacements that dissipate energy that would otherwise go into displacing the wall.\cite{Nakatani2003}

A key theoretical result of DW dynamics is the existence of two regimes of wall propagation separated by a critical field $H_{\text{c}}$.\cite{Malozemoff1979,Schryer1974} Drive fields $H$ below $H_{\text{c}}$ yield DW propagation that obeys a linear mobility relationship
\begin{equation}
 v(H)=\mu H \label{eq:1}
\end{equation}
where $v$ is the DW velocity and $\mu$ is the mobility. Drive fields $H>H_{\text{c}}$ produce a region of negative differential mobility followed by a second region of asymptotic linear mobility ($H\gg H_{\text{c}}$) with a significantly smaller mobility parameter than predicted (and observed) for $H<H_{\text{c}}$.

Domain wall motion can be studied by analytical models, which help provide physical insight into DW dynamics, and by more accurate and realistic numerical simulations. In the analytical one-dimensional (1D) model,\cite{Malozemoff1979,Schryer1974} a transverse DW has a mobility given by
\begin{subequations}
 \label{eq:2}
 \begin{eqnarray}
 \mu&=&\frac{\gamma \Delta}{\alpha}~~~~~~~~~~(H<H_{\text{W}}),\label{eq:2a}
 \\
 \mu&=&\frac{\alpha \gamma \overline{\Delta}}{1+{\alpha}^2}~~~~~~(H>H_{\text{W}}).\label{eq:2b}
 \end{eqnarray}
\end{subequations}
In Eq.~(\ref{eq:2}), $\gamma$ is the gyromagnetic ratio, $\Delta$ (or $\overline{\Delta}$) is the (average) DW width, and $\alpha$ is the Gilbert damping parameter. In the 1D model, the critical field is called the Walker field, $H_{\text{W}}$. The parameter $H_{\text{W}}$ specifies the onset of precessional spin motion that drives oscillatory DW motion and accounts for the reduced average velocity and lower mobility.  Numerical simulations of static and dynamic DW configurations in rectangular cross-section nanowire structures have shown that the spin configuration within a DW is governed by the nanowire cross-sectional parameters. Thin narrow wires support simple transverse wall (TW) spin structures whereas wider and thicker walls support more complex vortex wall (VW) structures.

Mobility measurements of DW propagation in Permalloy (Py) nanowires\cite{Beach2005} manifest the qualitative behavior predicted by the 1D model, but the measured $H_{\text{c}}$ is $\sim 10~\text{Oe}$, a factor of 10 lower than the 1D model value ($H_{\text{W}}\approx 100~\text{Oe}$); and the corresponding maximum velocity, $v(H_{\text{c}})$, is a factor of 10 lower than $v(H_{\text{W}})$. Numerical simulations of spin distributions within the propagating DWs account for this large discrepancy: the energy barrier for (anti) vortex formation is overcome at values of $H$ far below the onset of precessional motion given by the Walker field $H_{\text{W}}$. This (anti) vortex formation provides the mechanism for the onset of the low-mobility regime that limits DW propagation velocities to $v_{\text{c}}\sim 100~\text{m/s}$ at moderate applied fields. At sufficiently high longitudinal drive fields, the DW velocity can reach values equal to or exceeding the value at $H_{\text{c}}$.

Recent numerical simulation studies have explored possible ways to overcome velocity breakdown in nanowires by inhibiting (anti) vortex generation. The first\cite{Nakatani2003} of these predicted that edge roughness of an appropriate scale should disrupt antivortex formation at the wire edges. In a perfectly smooth wire, each nucleated antivortex core was observed to be gyrotropically driven into and across the wire, resulting in velocity breakdown behavior. When edge roughness on the scale of the antivortex core diameter ($\sim 10~\text{nm}$) was added to the simulations, the inhomogeneous local magnetostatic fields perturbed the nucleated antivortex cores, resulting in annihilations via spin-wave excitations before they were able to enter and slow the DW, thus eliminating the usual drop in velocity above $H_{\text{c}}$. However, the DW velocity remained self-limited to $v_{\text{c}}$ even at the fields exceeding $H_{\text{c}}$. As the DW speed approached $v_{\text{c}}$, an edge antivortex was generated and its subsequent decay dissipated energy, slowing the wall velocity.

A similar result was achieved in the simulations of Lee \textit{et al.}\cite{Lee2007apl} using a different means. The simulated mobility of a DW in an isolated Py nanowire was compared to the DW mobility in a Py nanowire lying atop a perpendicularly-magnetized underlayer. Above $H_{\text{c}}$, antivortex cores were nucleated at the edge of a propagating TW, but the stray field from the underlayer caused each core to be quickly expelled before it entered the wire. Again, however, the DW velocity increased with increasing field up to $H_{\text{c}}$, beyond which the wall velocity reached a plateau. Hence, while these studies suggest ways to partially suppress velocity breakdown, they did not predict an increase in the maximum velocity of DW propagation.

A possible route to enhancing the critical velocity was suggested in the numerical studies of Kunz \textit{et al.}\cite{Kunz2008} and Bryan \textit{et al.},\cite{Bryan2008} which examined the effects of a transverse in-plane bias field $H_{\text{bias}}$ on the dynamics of a TW. A related analytical study by Sobolev \textit{et al.}\cite{Sobolev1995} addressed the effects of transverse fields on the motion of (Bloch) DWs. A bias field, applied normal to wire axis, does not directly drive DW motion, but it does change the width $\Delta$ of the wall. When the field is aligned with the net wall moment, $\Delta$ increased; when it is anti parallel to the wall moment, $\Delta$ is decreased. An increase of $\Delta$ results in an enhancement of DW velocity, $v\propto \Delta$, as expected from the 1D model\cite{Malozemoff1979,Schryer1974} (Eq.~\ref{eq:2}). The velocity increase is modest, however, up to only $\sim 20\%$, and any benefit ceases beyond the breakdown drive-field threshold. In the oscillatory regime ($H>H_{\text{c}}$) Bryan \textit{et al.} found a region of enhanced DW velocity around $H=100~\text{Oe}$ for both positive and negative transverse bias field: the average DW velocity recovers to a value slightly above the peak value achieved at $H_\text{c}$. Glathe \textit{et al.}\cite{Glathe2008a,Glathe2008b} have reported corresponding enhancements in wider nanowire conduits.

In this paper, we describe the effects of a transverse bias field on nanowire guided DW dynamics in the case of wider wires that support more complex DW structures.\cite{McMichael1997,Nakatani2005,Lee2007,Yang2008prb,Hubert1998} We find strikingly different behavior from that predicted for wires of smaller cross section. While simulations show that high-mobility regions of DW propagation exist for $H>H_{\text{c}}$ in both thin (TW) and wider (VW) nanowire conduits, the wider conduits exhibit much higher transverse-field induced enhancement of DW velocity.

\section{\label{sec:2}Experiment}
A Py nanowire, $490~\text{nm}$ wide, $20~\text{nm}$ thick, and $35~\mu\text{m}$ long, was fabricated via focused ion beam milling [Fig.~\ref{fig:1}(a)]. One end incorporated a large magnetic pad that served as a nucleation and injection source of DWs; the other end was tapered into a point to inhibit DW nucleation. High-bandwidth magneto-optic polarimetry\cite{Nistor2006} was used to measure the time-dependent DW displacements under a combination of longitudinal drive field $H$ and transverse bias field $H_{\text{bias}}$ strengths. One advantage of using polarimetry to track DW displacements by time-of-flight techniques (and also instantaneous velocity by measuring the transient response\cite{Yang2008prb} as a DW sweeps across the polarimeter light spot) is that the mobility curves can be directly measured. A short injection pulse initiates DW propagation down the nanowire conduit and the velocity (both average and instantaneous at a prescribed point) can be measured for any combination of static longitudinal and transverse bias magnetic fields.

\section{\label{sec:3}Micromagnetic Simulations}
Micromagnetic simulation was used to understand the experimental results. The simulations were carried out using a version of LLG Micromagnetics Simulator developed by M. R. Scheinfein\cite{Scheinfein} that has been adapted to the University of Texas Lonestar Cluster\cite{Tacc} (a 5840 processor 64 bit Linux-based system capable of 62 TFLOP/sec peak performance). The numerical simulations were carried out using the accepted parameters for Permalloy ($\text{Ni}_{\text{80}}\text{Fe}_{\text{20}}$); saturation magnetization $M_{\text{s}}=800\times 10^3~\text{A/m}$, exchange constant $A=1.0\times 10^{-11}~\text{J/m}$, and Gilbert damping constant $\alpha = 0.01$. The unit cell size and integration time step were $4\times 4\times 20~\text{nm}^3$ and $0.3~\text{ps}$ respectively. To reduce computational time, moving boundary conditions were used, but selected moving boundary condition results were compared with corresponding fixed boundary condition results to ensure accuracy. The average DW velocity was obtained by averaging over many DW oscillation cycles (typically $100~\text{ns}$). The expanded capacity of the Dell-Cray Lonestar cluster (over single-processor systems) has allowed systematic exploration of nanowire constrained and guided DW properties (i.e., spin texture phase diagram and dynamics) over a wider parameter space and finer parameter increments than conveniently possible using single processor systems.

\section{\label{sec:3}Results and Discussion}
\begin{figure}
\includegraphics{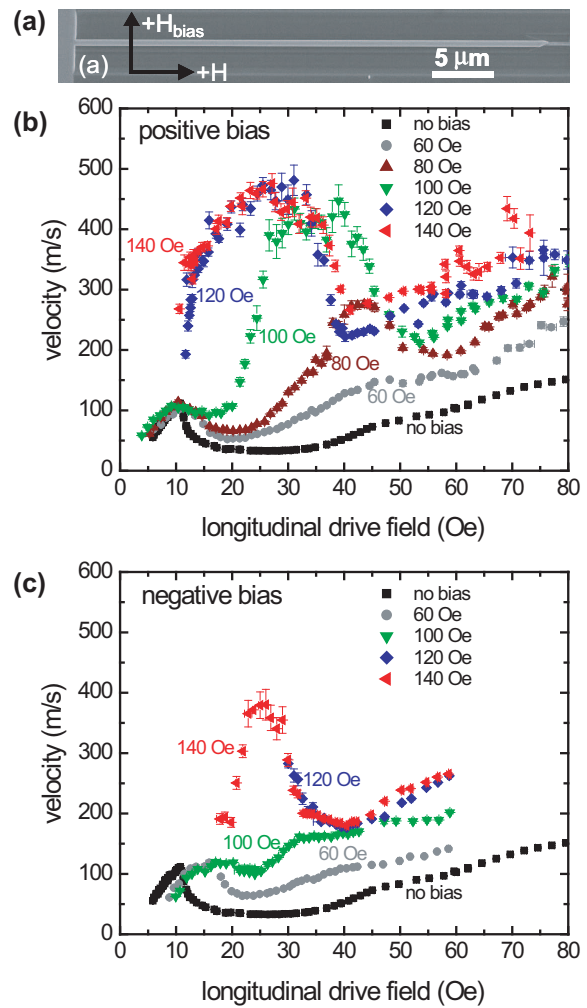}
\caption{\label{fig:1}(color online) Effect of transverse bias field on DW velocity. (a) Scanning electron micrograph of nanowire structure. (b) Average DW velocity versus $H$ for tail-to-tail wall under various positive bias fields. (c) Same as in (b), but with bias field direction reversed.}
\end{figure}

Figure~\ref{fig:1} shows the effects of $H_{\text{bias}}$ on the mobility curve $v(H)$ of a tail-to-tail DW in a $490~\text{nm}$ wide and $20~\text{nm}$ thick nanowire. Without $H_{\text{bias}}$, $v(H)$ exhibits typical velocity breakdown behavior\cite{Beach2005}: Wall velocity increases with increasing $H$, reaching a critical value $v_{\text{c}}=112~\text{m/s}$ at $H_{\text{c}}=10.6~\text{Oe}$. Just beyond this breakdown field, velocity drops abruptly as the wall undergoes oscillatory motion; at higher fields, $v$ begins to slowly increase again.

The application of $H_{\text{bias}}$ has relatively little effect on the mobility curve for $H<H_{\text{c}}$ and on the value of $H_{\text{c}}$ that marks the onset of oscillatory VW dynamics. However, $H_{\text{bias}}$ dramatically changes the velocity response above the breakdown transition. The family of curves $v(H)$ in Fig.~\ref{fig:1}(b) shows that the velocity with (positive) $H_{\text{bias}}$ is enhanced throughout the oscillatory regime. The enhancement is largest at intermediate values of $H$, where a peak feature develops. This peak grows and shifts to lower $H$ with increasing $H_{\text{bias}}$. At the largest $H_{\text{bias}}$ studied ($140~\text{Oe}$), the lower shoulder of the peak begins abruptly at $H_{\text{c}}$, just beyond which the velocity is seen to increase. The application of $H_{\text{bias}}$ transforms the character of the oscillatory transition from velocity breakdown to a velocity surge. Just above $H_{\text{c}}$, the DW velocity rapidly approaches $500~\text{m/s}$, nearly 5 times the (unbiased) critical velocity and 15 times larger than the velocity under the same drive field, without $H_{\text{bias}}$. The maximum mobility, $(\partial v/\partial H)_{\text{max}}$, is $\sim 30~\text{m/sOe}$, compared to the unbiased low-field ($H<H_{\text{c}}$ ) mobility of $11~\text{m/sOe}$. Similar behavior was observed for the (negative) $H_{\text{bias}}$ as shown in Fig.~\ref{fig:1}(c), as expected from the geometrical symmetry, although the mobility curves for $\pm H_{\text{bias}}$ are clearly not symmetric. These data represent an experimental indication that velocities far in excess of the critical velocity $v_{\text{c}}$ might be readily achieved.

\begin{figure*}
\includegraphics{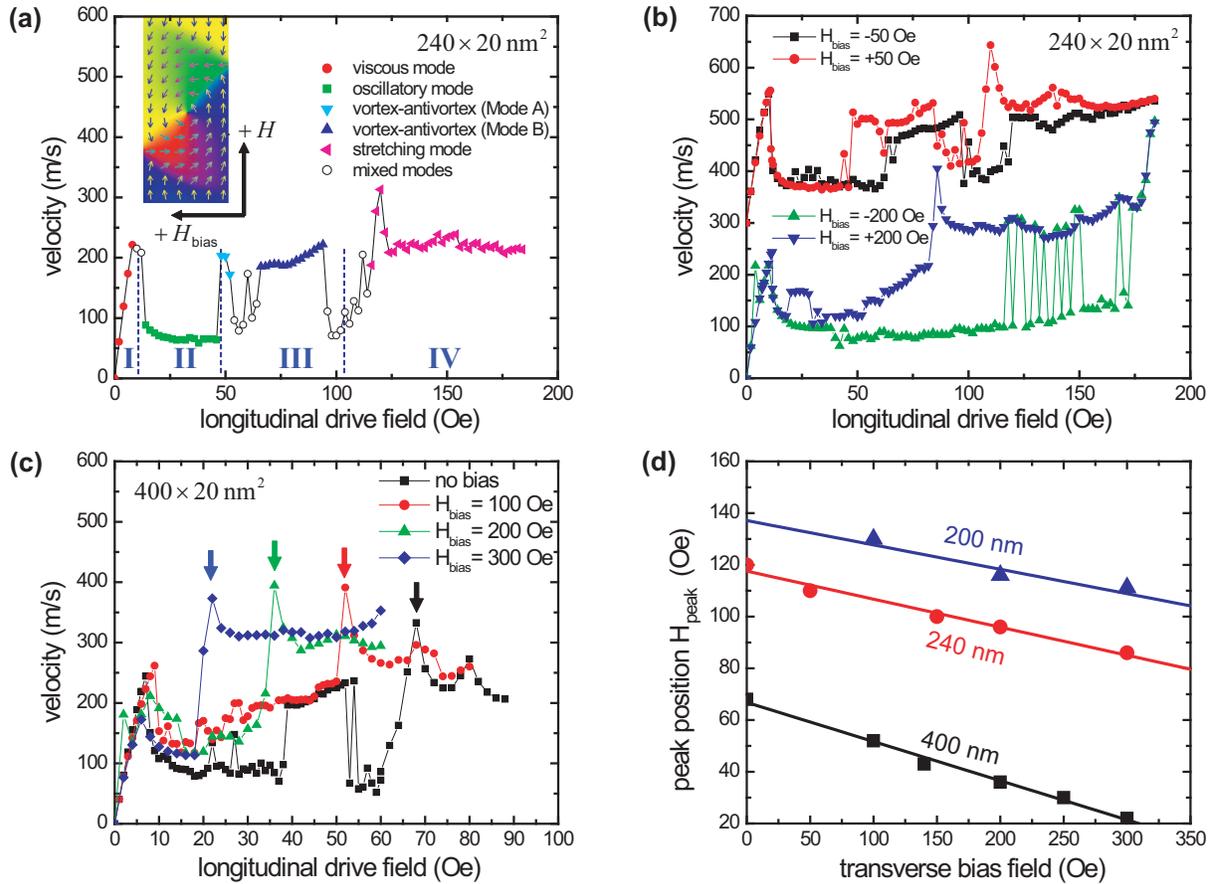}
\caption{\label{fig:2}(color online) Numerical simulations of bias-field-dependent mobility curves for vortex walls. (a) $H_{\text{bias}}=0$, (b) $H_{\text{bias}}=\pm 50~\text{Oe}$ and $\pm200~\text{Oe}$ for a $240~\text{nm}$ wide nanowire. Roman numerals and dashed lines designate regions of different dynamic behavior (refer to text). The inset [panel (a)] shows the counterclockwise-chirality vortex wall and the applied-field geometry in simulations. The curves for $H_{\text{bias}}=\pm50~\text{Oe}$ are vertically offset for clarity. (c) Same as (a) and (b) but for a $400~\text{nm}$ wide nanowire. Arrows indicate peaks in stretch mode (Region IV). (d) $H_{\text{peak}}$ as a function of bias field for different nanowire widths ($200~\text{nm}$, $240~\text{nm}$,  and $400~\text{nm}$).}
\end{figure*}

Figure~\ref{fig:2} displays the numerical simulations of the bias-field-dependent average DW velocity for VWs in $20~\text{nm}$ thick Permalloy nanowires with different widths as a function of $H$. In the simulations, the positive bias field was aligned parallel (anti-parallel) to the spin direction of the head (tail) portion of a counterclockwise-chirality ($c=1$) VW as shown in the inset of Fig.~\ref{fig:2}(a). The mobility curve $v(H)$ of a $240~\text{nm}$ wide nanowire for $H_{\text{bias}}=0$ in Fig.~\ref{fig:2}(a) exhibits regions of different dynamical behavior characterized by distinctly different spin distributions. Region I (viscous mode, $H<H_{\text{c}}$) corresponds to the stationary spin distributions (VW) that propagate with uniform velocity and linear mobility. Region II corresponds to complex oscillatory motion driven by precessional modes. Extended regions of continuous higher DW mobility at zero-bias field [Region III of Fig.~\ref{fig:2}(a)] have not been identified in prior simulations. Clear observation of this effect requires many calculations at closely-spaced drive-field increments, which is impractical with single-node versions of micromagnetic simulation codes. Region III is characterized by a stationary vortex-antivortex spin distribution that propagates uniformly at a significantly higher velocity than the oscillatory motion modes. Region IV is characterized by a stretching mode that contains a vortex (slow portion) and topological edge defect (fast portion) structure that propagates at a relatively high average velocity.

Note [from Fig.~\ref{fig:2}(b)] that positive bias fields shift the onset of the (no-bias) vortex-antivortex resonance (Region III), as well as the onset of the stretching mode (Region IV) to lower values of $H$. Negative bias fields, however, suppress the additional dynamic modes (vortex-antivortex resonance mode and stretching mode), as shown in the curve for $H_{\text{bias}}=-50~\text{Oe}$ in Fig.~\ref{fig:2}(b). For a higher negative bias field ($H_{\text{bias}}=-200~\text{Oe}$), the additional dynamic modes are totally suppressed for the driving fields of $H<120~\text{Oe}$. Beyond this field ($H>120~\text{Oe}$), the vortex chirality starts to change its sign [from counterclockwise ($c=1$) to clockwise ($c=-1$)] stochastically and the chirality switching probability reaches almost $100\%$ for $H>170~\text{Oe}$. Once the chirality switching occurs, the counterclockwise vortex with a negative bias field has the same dynamics as the clockwise vortex with a positive bias field, as shown in the curves for $H_{\text{bias}}=\pm200~\text{Oe}$ in Fig.~\ref{fig:2}(b).

Figure~\ref{fig:2}(c) extends the study of simulated mobility curves to higher (positive) $H_{\text{bias}}$ for a wider ($400~\text{nm}$) nanowire structure. The effect of $H_{\text{bias}}$ [as seen in Fig.~\ref{fig:2}(b)] is to continue to shift the onset of Region III and Region IV to lower values of $H$. At $H_{\text{bias}}=300~\text{Oe}$, the onset of Region IV (stretching mode) is shifted to below $H=20~\text{Oe}$. The high-mobility stretching mode is stabilized by $H_{\text{bias}}$ throughout most of the low-mobility Region III. Note the progression of $H_{\text{bias}}$-driven peak position ($H_{\text{peak}}$) of the stretch mode indicated by arrows in Fig.~\ref{fig:2}(c). $H_{\text{peak}}$ depends on the transverse bias fields and the widths of nanowires, as shown in Fig.~\ref{fig:2}(d). We attribute the dramatic increase of DW velocity and the shifting of the arrow-designated peak in Fig.~\ref{fig:2}(c) associated with the $H_{\text{bias}}$ to stabilization of the stretching mode identified in the numerical simulations. The experimental mobility curves (Fig.~\ref{fig:1}) exhibit the same general behavior as $H_{\text{bias}}$ is increased. Note that the chirality of injected domain wall in the experiment (Fig.~\ref{fig:1}) depends on the direction of transverse bias field. The spin direction of the head (tail) portion of injected DW is parallel (anti-parallel) to the applied transverse bias field, which corresponds to the positive bias field configuration in the simulations [see the inset of Fig.~\ref{fig:2}(a)].

\begin{figure}
\includegraphics{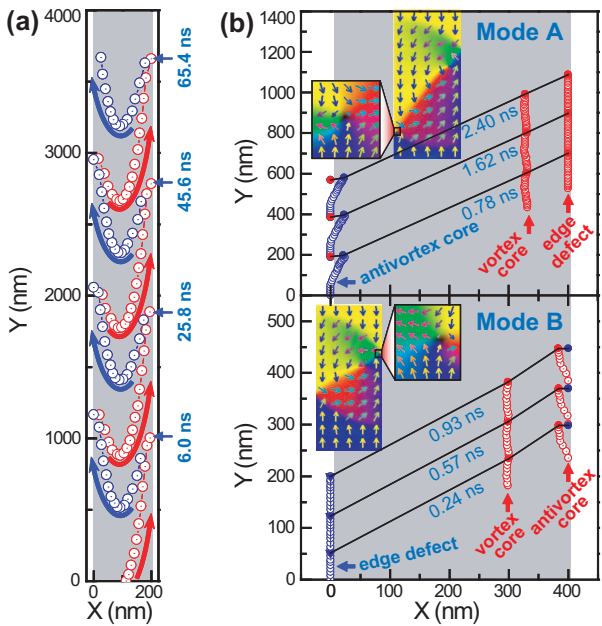}
\caption{\label{fig:3}(color online) Numerical simulations of DW spin distributions and trajectories. (a) Trajectory of a vortex core (in a $200~\text{nm}$ wide wire, $H=25~\text{Oe}$ and $H_{\text{bias}}=0$) showing core polarity reversal at wire edges accompanied by velocity oscillation (Region II). (b) Trajectories of topological defects (edge defect, vortex core, and antivortex core) in a $400~\text{nm}$ wide wire, corresponding to vortex-antivortex resonance mode (Region III). Insets show spin distributions with magnified antivortices. Antivortices are periodically generated at the left edge in Mode A ($H=27~\text{Oe}$, $H_{\text{bias}}=0$); at the right edge in Mode B ($H=47~\text{Oe}$, $H_{\text{bias}}=0$). Colors of symbols in (a) and (b) represent polarities of topological defects: red (+), blue (-).}
\end{figure}

Figure~\ref{fig:3} provides additional insight into the DW spin distributions associated with specific regions of the mobility curve shown in Fig.~\ref{fig:2}(a). The uniformly propagating stationary mode (Region I, $H<H_{\text{c}}$) is well understood,\cite{Malozemoff1979,Schryer1974,Hubert1998} (Walker solution). The complex oscillatory motion for $H_{\text{c}}$ (Region II) has been explored numerically\cite{Nakatani2005,Lee2007} and experimentally\cite{Yang2008prb}. The simulated displacement of an oscillatory mode [illustrated as a displacement record in Fig.~\ref{fig:3}(a)] is a particularly simple example that occurs in a narrow (200 nm) wire where mode hopping is suppressed by geometrical constraints. This panel illustrates the trajectory of a vortex core as it sweeps from edge-to-edge, changing polarity of vortex core at each edge, while conserving the vortex chirality. The DW experiences temporary reversal of longitudinal velocity after polarity reversal at the edges. The low average velocity (for $H>H_{\text{c}}$) results from the oscillatory motion.

Fig.~\ref{fig:3}(b) illustrates the dynamics of the vortex-antivortex resonance modes that account for the region of increased mobility in the simulations (Region III in Fig.~\ref{fig:2} with $H_{\text{bias}}=0$). Two topologically distinct vortex-antivortex resonance modes, Mode A and Mode B in Fig.~\ref{fig:3}(b), are identified, and both propagate at high velocities due to suppression of the oscillatory behavior illustrated in Fig.~\ref{fig:3}(a). The spin distributions of the two modes (shown as insets) consist of a vortex and an antivortex near one nanowire edge. In both cases, the antivortices stabilize the position of vortex cores, so that the spin distribution propagates as a nearly stationary state at high velocity. Figure~\ref{fig:3}(b) shows minor oscillatory (time-dependent transverse displacements) of the (left) antivortex core (Mode A) and corresponding transverse displacements of the (right) antivortex core (Mode B). However, the primary dynamics are characterized by uniform stationary motion along the wire. This result [Mode A and B of Fig.~\ref{fig:3}(c)] demonstrates that it is possible to stabilize (at $H_{\text{bias}}=0$) stationary high-mobility spin distributions above $H_{\text{c}}$. We have observed some evidence of this behavior in our mobility measurements (note the structure in the $H_{\text{bias}}=0$ mobility curve in Fig.~\ref{fig:1}, and similar effects in prior reported mobility curves\cite{Beach2006}).

\begin{figure}
\includegraphics{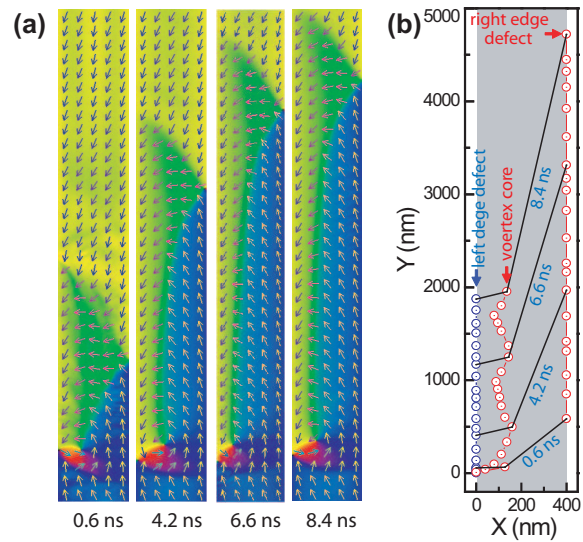}
\caption{\label{fig:4}(color online) Numerical simulations of DW spin distributions and trajectories of the stretch mode (Region IV). (a) Time evolution of spin distributions, (b) corresponding trajectories (in a $400~\text{nm}$ wide wire, $H=40~\text{Oe}$ and $H_{\text{bias}}=200~\text{Oe}$). The (right) edge defect propagates much faster than the (left) vortex-antivortex structure. Color of symbols in (b) represents polarities of topological defects: red (+), blue (-).}
\end{figure}

\begin{figure*}
\includegraphics{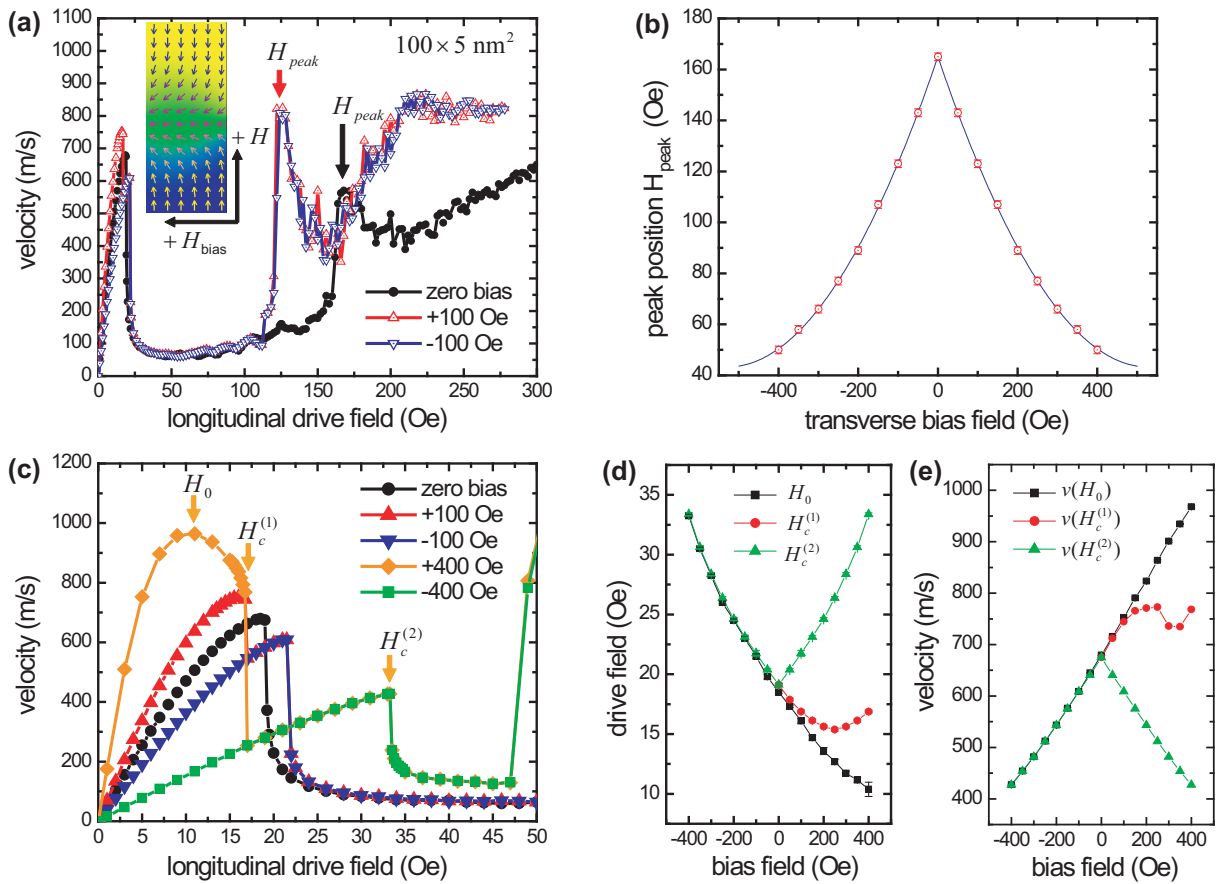}
\caption{\label{fig:5}(color online) Numerical simulations of bias-field-dependent mobility curves for a transverse wall (nanowire dimensions:$100~\text{nm}$ wide $5~\text{nm}$ thick). (a) $H_{\text{bias}}=0$ and $H_{\text{bias}}=\pm100~\text{Oe}$. Arrows indicate the peak position, $H_{\text{peak}}$, for $H>H_{\text{c}}$. Inset, transverse wall and applied field geometry in simulation. (b) Peak position, $H_{\text{peak}}$, as a function of bias field. (c) $H_{\text{bias}}=0$, $\pm100~\text{Oe}$, and $\pm400~\text{Oe}$. The curves for positive bias fields show two-step breakdown behavior ($H^{\text{(1)}} _{\text{c}}$ and $H^{\text{(2)}} _{\text{c}}$) and have a maximum viscous velocity at $H_{\text{0}}$. (d) Bias-field dependence of $H_{\text{0}}$, $H^{\text{(1)}} _{\text{c}}$, and $H^{\text{(2)}} _{\text{c}}$, (e)corresponding velocities.}
\end{figure*}

Figure~\ref{fig:4} illustrates the spin distributions and dynamics of the stretching mode that accounts for the $H_{\text{bias}}$ induced high-mobility behavior shown in experiments (Fig.~\ref{fig:1}) and simulations (Fig.~\ref{fig:2}). The spin distribution [refer to Fig.~\ref{fig:4}(a)] associated with the peak in mobility in Region IV consists of a vortex-antivortex structure near the left edge of the wire and an edge defect pinned at the opposite edge. The spin distribution of the stretch mode is similar to the voretex-antivortex mode (Region III). Periodically generated antivortices from the left edge stabilize the vortex core position, but the dynamics are different. The vortex-antivortex structure in the stretch mode propagates at a different velocity than the (right) edge defect (compare the time-labeled trajectories in Figs.~\ref{fig:3} and \ref{fig:4}). The resulting motion is characterized by a stretching DW with modulated width but with basically a stationary spin distribution. An applied transverse bias field stabilizes the stretching mode over a wide range of drive field strengths (illustrated by arrows in Fig.~\ref{fig:2}). The mobility of the stretch mode is about four times the mobility of oscillatory modes for drive fields above the critical field.

In order to explore the possible connection between bias-field induced enhancement of DW velocity in narrow and wide nanowire conduits, we carried out numerical simulations of the mobility curves for the same narrow nanowire geometry reported by Bryan \textit{et al.}\cite{Bryan2008} Figure~\ref{fig:5}(a) displays the simulated mobility curve for a $100~\text{nm}$ wide $5~\text{nm}$ thick Py nanowire (over a broader drive field range and using smaller field increments using the same parameters used in the prior simulation). The positive transverse bias fields were aligned parallel to the domain wall magnetization of a TW with counterclockwise chirality along the y-axis, as shown in the inset of Fig.~\ref{fig:5}(a). Without a bias field, a TW propagates viscously for $H<H_{\text{c}}$. For $H>H_{\text{c}}$, the chirality of TW changes its sign periodically from counterclockwise to clockwise via anti-vortex generation, resulting in DW oscillation. The qualitative behavior of the simulated $H_{\text{bias}}=0$ mobility curve for the narrow-wire structure is similar to the corresponding simulation result [Fig.~\ref{fig:2}(a)] for the wider nanowire. Both simulations manifest linear high mobility regions below $H_{\text{c}}$, and a lower-mobility region following the negative differential mobility region just above $H_{\text{c}}$; and both simulated mobility curves manifest narrow regions of enhanced mobility above $H_{\text{c}}$ characterized by a rapid onset of high differential mobility. The onset of the high mobility region for the narrow wire structure occurs at $H=165~\text{Oe}$ (considerably above the corresponding threshold for the $240~\text{nm}$ and $400~\text{nm}$ structures which occur around $70~\text{Oe}$ and $120~\text{Oe}$, respectively). This high-mobility zero bias field mode exists at drive field strength beyond the range studied by Bryan \textit{et al.}

The peak shifting behavior by a transverse bias field is similar to the case of VW, as indicated by arrows in Fig.~\ref{fig:5}(a). However, different from the case of VW, the peak position shift for TW as a function of transverse bias field [Fig.~\ref{fig:5}(b)] is symmetric and unidirectional. For $H>H_{\text{c}}$, the mobility curves for $+H_{\text{bias}}$ and $-H_{\text{bias}}$ become identical due to the the periodic chirality switching as shown from the mobility curves for $H_{\text{bias}}=\pm100~\text{Oe}$. Hence, both positive and negative transverse bias fields shift the peak position to lower axial driving fields.

For a small driving field ($H<H_{\text{c}}$), the increased (decreased) DW width by positive (negative) transverse bias fields results in the increased (decreased) DW mobility [Eq.~(\ref{eq:2a})], as shown in Fig.~\ref{fig:5}(c). Furthermore, the application of transverse bias fields induces asymmetry of the anti-vortex (winding number $n=-1$) emission energy barrier from the left and right edges of nanowires. Due to the asymmetric energy barrier, two-step breakdown ($H^{\text{(1)}} _{\text{c}}$ and $H^{\text{(2)}} _{\text{c}}$)\cite{Seo2010} can be observed for positive transverse bias fields, as indicated by arrows in Fig.~\ref{fig:5}(c). For $H_{\text{bias}}=400~\text{Oe}$, DW velocity increases with increasing drive fields, reaches it's maximum viscous velocity [$v(H_{\text{0}})=967~\text{m/s}$], and then decreases until the first breakdown occurs at $H^{\text{(1)}} _{\text{c}}$. This negative differential mobility in the viscous mode is similar to the analytical results of the transverse bias field effect on the Bloch wall velocity, studied by Sobolev \textit{et al.}\cite{Sobolev1995} In the process of the first breakdown, the TW chirality changes from counterclockwise to clockwise via an anti-vortex ($n=-1$) emission from the topological right-side edge defect ($n=-1/2$), so the the bias field aligns anti-parallel to the net transverse magnetic moment of the clockwise TW. Therefore, beyond $H^{\text{(1)}} _{\text{c}}$, DW velocity is independent of the sign of transverse bias fields. Figure~\ref{fig:5}(d) shows the bias-field dependence of $H_{\text{0}}$, $H^{\text{(1)}} _{\text{c}}$, and $H^{\text{(2)}} _{\text{c}}$. The corresponding velocities are displayed in Fig.~\ref{fig:5}(d). Note that, different from the case of TW, a VW has no net transverse magnetic moment. Therefore, the transverse bias fields have little effect on the VW velocity for $H<H_{\text{c}}$.

\section{\label{sec:3}Summary and Conclusion}
In summary, these results show that application of a static transverse magnetic field can dramatically enhance the velocity of DW propagation in magnetic nanowires. Moderate bias fields have little effect in the low-field regime or on the onset of drive-field-induced precession. However, the DW dynamics in the precessional regime is changed dramatically. Micromagnetic simulations suggest that several internal modes are excited within the DW, and that these modes can couple such that the DW structure becomes dynamically stable. The dynamically stable modes propagate as a stationary spin distribution (vortex-antivortex resonance mode), or as a \textquotedblleft\textit{modulated}\textquotedblright  stationary spin distribution (stretch mode). A transverse field can enhance these effects and extend the region of high-speed propagation in the case of the stretch mode. This is borne out qualitatively in the experiments, wherein the velocity surges just beyond the critical field for vortex nucleation. The character of the mobility curve in the precessional regime depends nontrivially on the field polarities and, presumably, vortex chirality. Numerical simulations for narrow nanowire geometry that supports TW spin distributions reveal zero-bias higher mobility spin distributions above $H_{\text{c}}$ that shift to lower values of drive field when transverse bias fields are applied (similar to the wider nanowire effect).

\begin{acknowledgments}
This work was supported by the NSF under Grant $\#$DMR-0404252 and DMR-0903812. Nanowires were fabricated using facilities of the Center for Nano and Molecular Science and Technology at UT-Austin. Simulations were conducted using HPC resources of TACC at UT-Austin.
\end{acknowledgments}

\end{document}